\documentclass{desyproc}

\begin{document}
\title{Hidden photon CDM search at Tokyo}

\author{{\slshape  Jun'ya Suzuki, Yoshizumi Inoue, Tomoki Horie, Makoto Minowa}\\[1ex]
The University of Tokyo, Japan}

\contribID{familyname\_firstname}

\confID{11832}  
\desyproc{DESY-PROC-2015-02}
\acronym{Patras 2015} 
\doi  

\maketitle

\begin{abstract}
We report on a search for hidden photon cold dark matter (HP CDM) using a novel technique with a dish antenna. 
We constructed two independent apparatus: one is aiming at the detection of the HP with a mass of $\sim\,\rm{eV}$ which employs optical instruments, 
and the other is for a mass of $\sim5\times10^{-5}\,\rm{eV}$ utilizing a commercially available parabolic antenna facing on a plane reflector. 
From the result of the measurements, we found no evidence for the existence of HP CDM and set upper limits on the photon-HP mixing parameter $\chi$. 
\end{abstract}

\section{Introduction}

Astronomical observations of the past decades reveal that 
there exists invisible non-baryonic matter (dark matter, DM) in the universe. 
Exploring the nature of DM is one of the most important issues in astrophysics and cosmology today, 
and a variety of experiments have been carried out to directly detect DM particles. 

The most prominent candidate for DM is the Weakly Interacting Massive Particle (WIMP), 
and most of the current experiments aim at detection of WIMPs. 
However, there are alternative candidates to account for the features of DM, 
and Weakly Interacting Slim Particles (WISP), e.g. axion-like particles (ALP)  or hidden-sector photons (HP), can be the main component of DM~\cite{Arias}. 


Hidden photon CDM can be experimentally investigated 
via kinetic mixing $(\chi/2)F_{\mu\nu}\tilde{X}^{\mu\nu}$ between photons and hidden photons. 
For example, the Axion Dark Matter eXperiment (ADMX)~\cite{ADMX_detector}, 
which employs a resonant cavity and magnetic field to search for axion dark matter, 
also has sensitivity to hidden photon CDM, 
and its non-detection of the signal~\cite{ADMX_result1, ADMX_result2, ADMX_result3, ADMX_result4, ADMX_result5} was translated to the upper limit for the kinetic mixing parameter $\chi$~\cite{Arias}. 

Additionally, a novel method with a spherical mirror to search for HP CDM 
was recently proposed~\cite{Horns}, with which 
wider mass-range can be probed without rearranging the set-up. 
In this method, 
ordinary photons of energy $\omega \simeq m_{\gamma '}$ induced by HP CDM via kinetic mixing 
are emitted in the direction perpendicular to the surface of the mirror, 
resulting in concentration of the power to the center of the mirror sphere. 

This method using a spherical reflector is extremely simple,  
and can be implemented relatively easily. 
To confirm its feasibility in real situations, we planned and carried out two experiments to search for HP CDM in two different mass regions: one is for $m_{\gamma'}\sim\rm{eV}$ using optical equipments and the other for $m_{\gamma'}\sim50\,\mu\rm{eV}$ employing RF instruments. 
Here we report on the preparations and the results of those searches for HP CDM using the dish method. 

\section{Optical search}

\begin{wrapfigure}{r}{0.4\textwidth}
\includegraphics[width=0.4\textwidth]{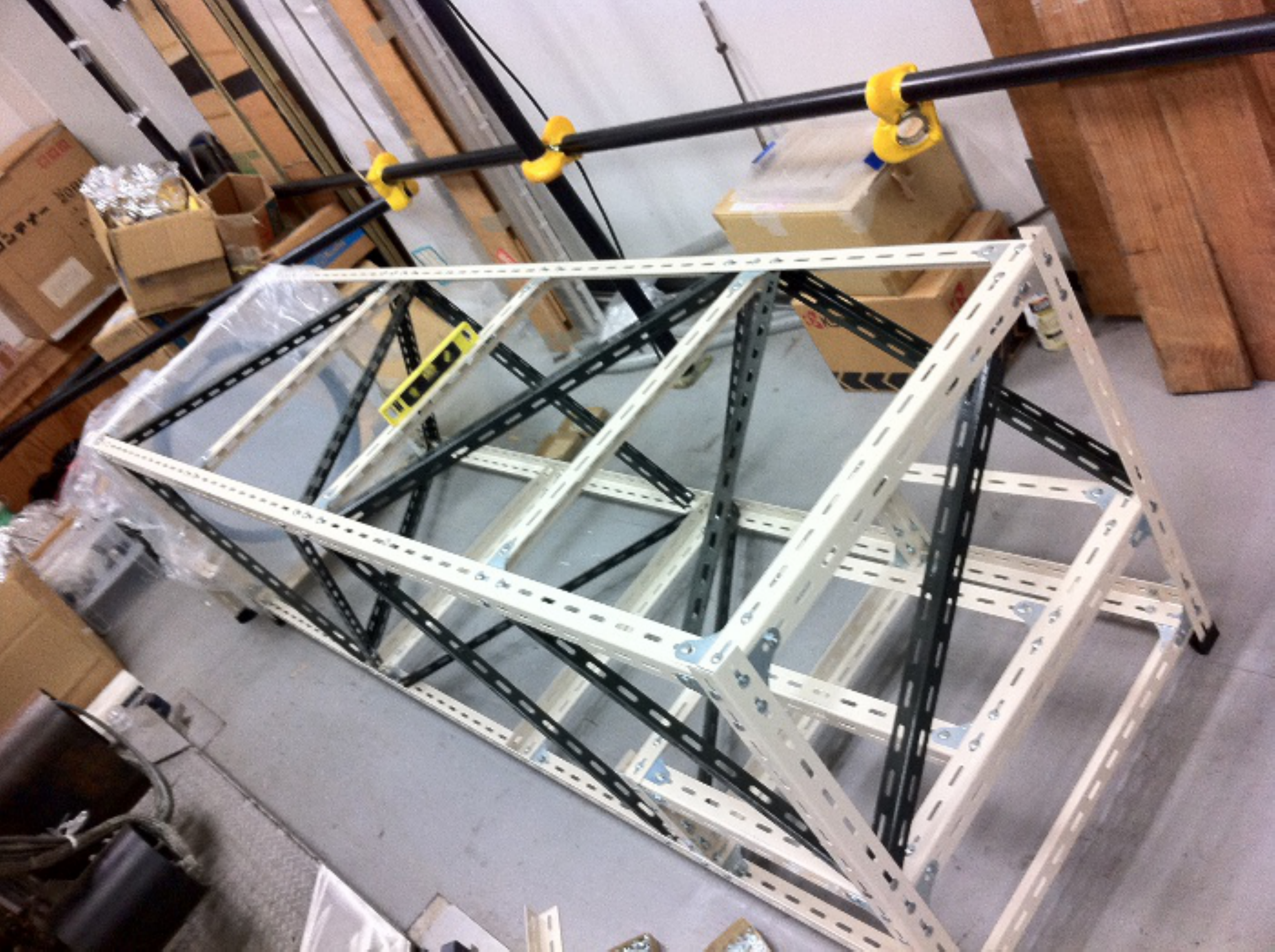}
\caption{The set-up for the optical seach. After installing optical equipments, this frame was wrapped with black polyethylene sheets to block ambient light. }\label{fig:opt_pict}
\end{wrapfigure}

For the seach in $m_{\gamma'}\sim \rm{eV}$, we need a spherical mirror and a photodetector.
Non-relativistic HPs near the surface of a reflector 
induce emission of photons in the direction perpendicular to the surface. 
A photodetector is placed at the point of convergence and detect emitted photons. 

We used a parabolic mirror  as a `dish'. 
The parabolic mirror is 500mm in diameter, 1007mm focal length and the focal spot diameter is 1.5mm. 
We used a parabolic surface instead of a spherical surface originally proposed in Ref.~\cite{Horns}
to reuse the mirror which had been employed in the solar HP helioscope~\cite{Mizumoto}. 
From the diameter and the focal length of the parabolic mirror, photons emitted perpendicularly to the surface 
are calculated to concentrate to a small area of 4 mm in diameter at twice the focal length of the mirror, 
which is small enough compared to the effective area of the photodetector. 

A photomultiplier tube (PMT) was employed as the detector of emitted photons. 
We selected Hamamatsu Photonics R3550P because of its low dark count rate of $\sim 5 \rm{Hz}$. 
We used a motorized stage to shift the position of the PMT, which enabled us to measure background noise. 

The mirror and the detector were mounted on a steel frame, which rigidly holds the arrangement (Fig.~\ref{fig:opt_pict}). 
After installing optical equipments, this frame was wrapped with black polyethylene sheets to shield from ambient light. 
Additionally, whole the set-up was installed in a light-tight box of $1\rm{m}\times1\rm{m}\times3\rm{m}$ to attain higher light-tightness. 

With this set-up, we carried out the experimental search for HP CDM in the eV mass range~\cite{jsuzuki}. 
The overall duration of the measurement was $8.3\times10^5 \,\rm{s}$ for each with the PMT 
at the position of convergence of the HP CDM signal (signal, S)
and at the position displaced by 25 mm from position S (background, B). 
We found no excess in count rate measured at position S compared to at position B. 
We translated this non-detection result to the limit for the mixing parameter $\chi$ (Fig.~\ref{fig:result}).

\section{RF search}

We also targeted detection in $\rm{K_{u}}$ band ($\sim 12\,\rm{GHz}$) for the feasibility test of the `dish' method. 
We can use commercially available dish antennas for this frequency region, 
though they usually have parabolic shape, which cannot be approximated as spherical shape 
because of their short focal lengths compared to their diameters. 
In order to overcome this problem, 
we let our dish face a plane reflector, 
from which plane radio wave of HP CDM origin would be emitted perpendicularly to the surface. 
Because parabolic dishes concentrate plane wave to their focal point, 
the amplification of HP CDM signal properly works. 

We used Anstellar SXT-220 as a dish, which is 2.2 m diameter and designed for CS broadcast reception. 
A huge plane reflector was constructed by combining four alminum plates on a rigid frame. 
For the converter, we selected Norsat 4506B, which down-converts the signal with the local frequency of 11 GHz. 

\begin{wrapfigure}{r}{0.45\textwidth}
\includegraphics[width=0.2\textwidth]{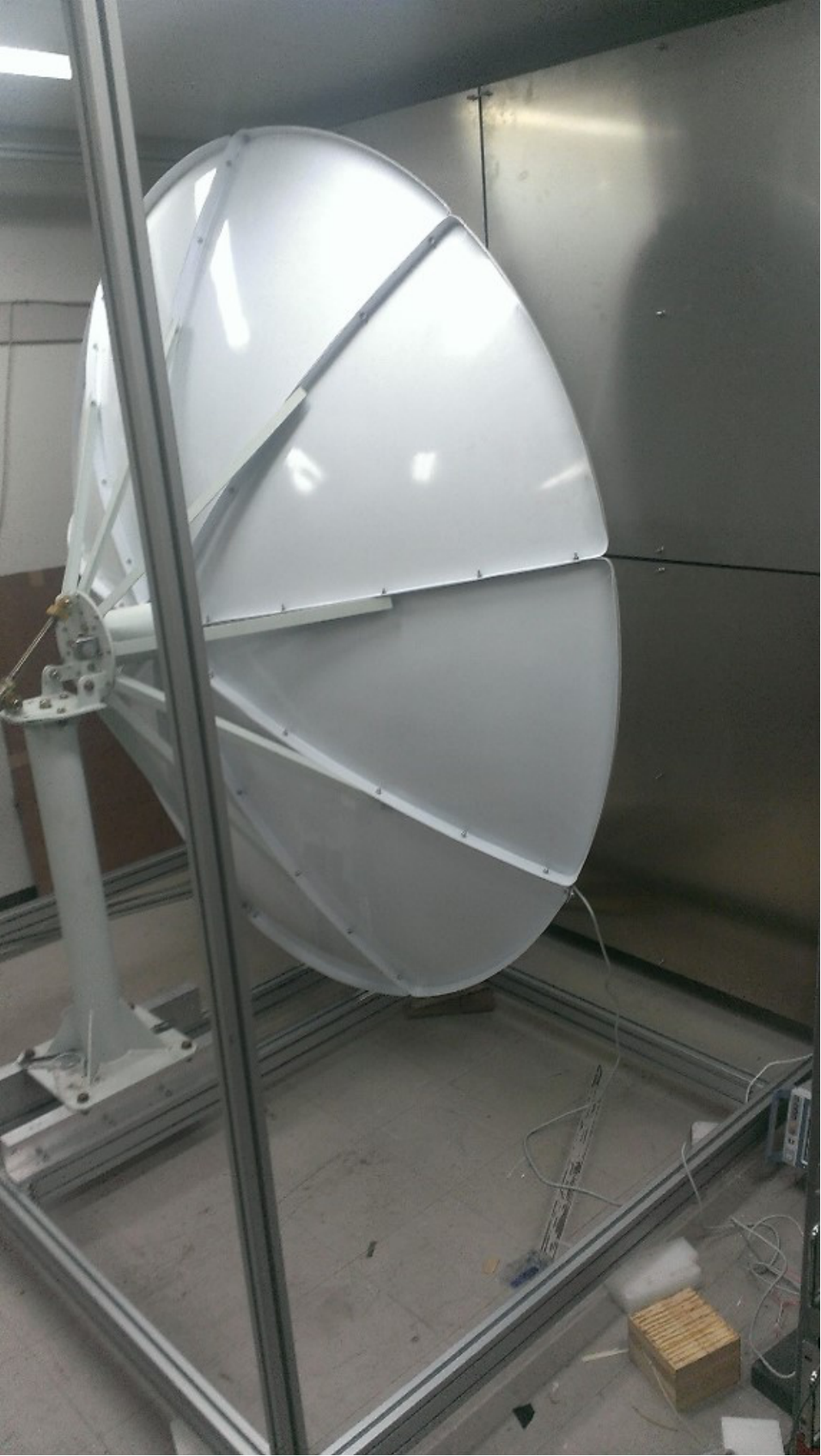}
\includegraphics[width=0.2\textwidth]{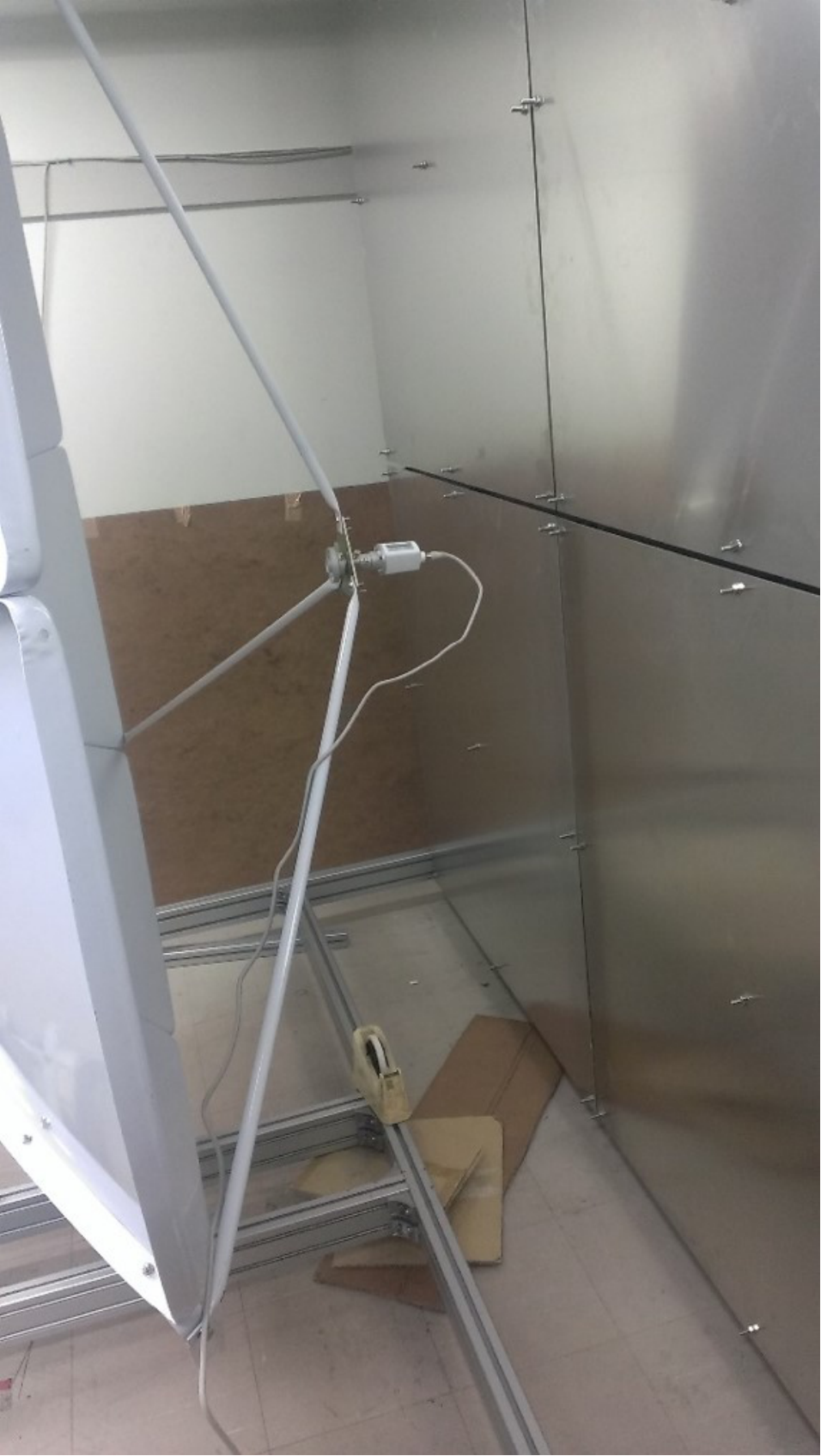}
\caption{The set-up for the search in $\rm{K_{u}}$ band. The parabolic dish designed for CS broadcast reception faces on the plane reflector made up of four alminum plates. }
\label{fig:rf_pict}
\end{wrapfigure}

The output of the converter was connected to the FFT analyzer, Rohde \& Schwarz FSV-4. 
The signal of the existence of HP CDM would be seen as a spectral line with a broadening of $\Delta f/f\sim10^{-6}$ due to the velocity dispersion of DM. 

After the calibration, the set-up for the experimetal search was constructed 
by setting the dish in front of the plane reflector (Fig.~\ref{fig:rf_pict}). 

\begin{figure}
\begin{center}
\includegraphics[width=0.8\textwidth]{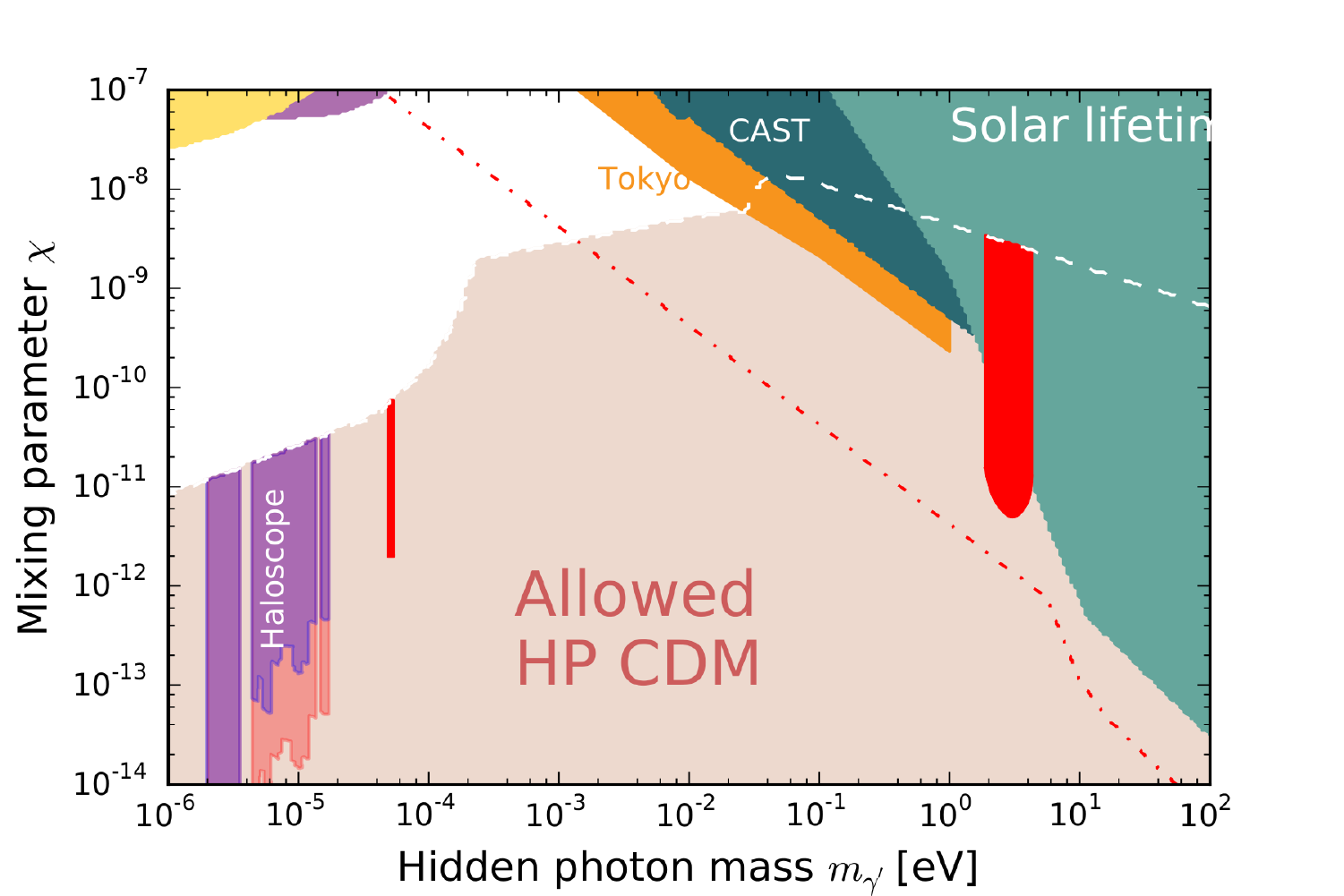}
\caption{Preliminary results of the experimental searches for HP CDM. 
The vertical axis shows the mixing parameter $\chi$, 
and the horizontal axis shows the mass of hidden-photon $m_{\gamma'}$. 
The red colored regions are excluded by our results for two experimental set-ups.
With the optical set-up, we excluded the area around $m_{\gamma'}\sim\rm{eV}$. 
The search in $\rm{K_u}$ band excluded the region around $m_{\gamma'}\sim50\,\mu\rm{eV}$. 
For descriptions of other colored areas, see Ref.~\cite{jsuzuki}. 
}
\label{fig:result}
\end{center}
\end{figure}

Using this set-up, we actually carried out the experimental search for four days. 
We observed no signal-like excess in the power spectrum and set an upper limit for the parameter $\chi$ (Fig.~\ref{fig:result}). 
Although the limit is narrow in the sensitive mass region, 
we can expand it only by replacing the converter 
for the one which is capable of handling wider frequency range. 

\section{Conclusion}

We constructed two apparatus utilizing a novel method using a dish antenna. 
One uses an optical mirror for the survey in $m_{\gamma'}\sim\rm{eV}$, 
and the other uses a dish antenna for CS broadcast reception to search HPs with $m_{\gamma'}\sim50\,\mu\rm{eV}$. 
We actually carried out the experimetal search, and found no evidence for the existence of HP CDM.
From the result, we set upper limits on the photon-HP mixing parameter $\chi$ in two different mass regions (Fig.~\ref{fig:result}).

\section*{Acknowledgments}
T. Horie acknowledges support by Advanced Leading Graduate Course for Photon Science (ALPS) at the University of Tokyo. 
This reaserch is supported by the Grant-in-Aid for challenging Exploratory Research by MEXT, Japan,  and also by the Research Center for the Early Universe, School of Science, the University of Tokyo.
 

\begin{footnotesize}

\end{footnotesize}


\end{document}